\documentclass[runningheads, envcountsect]{llncs}

\makeatletter
\renewcommand\subsubsection{\@startsection{subsubsection}{3}{\z@}%
	{-18\p@ \@plus -4\p@ \@minus -4\p@}%
	{0.5em \@plus 0.22em \@minus 0.1em}%
	{\normalfont\normalsize\bfseries\boldmath}}
\makeatother
\usepackage{graphicx}
\usepackage{here}
\usepackage{amsfonts}
\usepackage[hyphens]{url}
\usepackage{algorithm}
\usepackage[noend]{algpseudocode}

\usepackage{amsthm}
\usepackage{amssymb}
\usepackage{listings}
\usepackage{multirow}
\usepackage{mathtools}
\usepackage{array}
\theoremstyle{definition}
\newtheorem{defn}{Definition}[section]
\newcommand{\QIFone}{\mbox{QIF}_1}
\newcommand{\QIFtwo}{\mbox{QIF}_2}
\newcommand{\Pre}{\mbox{pre}}
\newcommand{\pre}{\mbox{pre}}
\newcommand{\Post}{\mbox{post}}
\newcommand{\QIF}{\mbox{QIF}}

\newcommand{\True}{\top}
\newcommand{\False}{\bot}

\begin{document}
\setcounter{secnumdepth}{3}
%
%\title{Contribution Title\thanks{Supported by organization x.}}
%\title{A compositional method for bounding dynamic leakage}
\title{On the Compositionality of Dynamic Leakage and Its Application to the Quantification Problem}
\titlerunning{Compositionality of Dynamic Leakage and Its Application}
\author{Bao Trung Chu \and
Kenji Hashimoto \and
Hiroyuki Seki}
\authorrunning{B.\ T.\ Chu et al.}
\institute{Graduate School of Information Science, Nagoya University, Japan}
\maketitle              % typeset the header of the contribution
\begin{abstract}

%The abstract should briefly summarize the contents of the paper in
%Quantitative information flow (QIF) was proposed as a refinement of
%non-interference property in security assurance.
Quantitative information flow (QIF) is traditionally defined 
as the expected value of information leakage over all feasible program runs
and it fails to identify vulnerable programs where only limited number of runs
leak large amount of information. 
As discussed in Bielova (2016), a good notion for dynamic leakage
and an efficient way of computing the leakage are needed. 
To address this problem, the authors have already proposed two notions 
for dynamic leakage 
%the authors prove the hardness of computing the quantity in general, 
and a method of quantifying dynamic leakage based on model counting. 
%Though it works relatively efficiently on small QIF benchmarks, 
%it is difficult to scale up closer to the size of real-world systems. 
Inspired by the work of Kawamoto et. al. (2017), 
this paper proposes two efficient methods for computing dynamic leakage, 
a compositional method along with the sequential structure of a program 
and a parallel computation based on the value domain decomposition. 
For the former, we also investigate both exact and approximated calculations. 
From the perspective of implementation, 
we utilize binary decision diagrams (BDDs) and deterministic decomposable negation 
normal forms (d-DNNFs) to represent Boolean formulas in model counting. 
Finally, we show experimental results on several examples.

\keywords{Dynamic leakage  \and Composition \and Quantitative Information Flow \and BDD \and d-DNNF.}
\end{abstract}
\section{Introduction}
Since first coined by \cite{GM82} in 1982, noninterference property has become 
the main criterion for software security. 
A program is said to satisfy noninterference 
if any change in confidential information does not affect a publicly observable output of that program. 
However, noninterference is so strict that it blocks many useful, 
yet practically safe systems and protocols, such as password checkers, anonymous voting protocols, recommendation
systems and so forth. Quantitative information flow (QIF) was introduced to loosen security 
criterion in the sense that, instead of seeking \textit{if there is} a case that a confidential input affects a 
public output, computing \textit{how large} that effect is. That is, if QIF of a program is insignificant, 
the program is still judged as secure. Because of its flexibility, QIF gains much attention in recent years. 
%from the research community. 
But it has an inherent shortcoming as shown in the example below.
\begin{example}
	\label{ex:1.1}
	Consider the following program taken from \cite{CHS19}.
	\begin{quote}
		if $source<16$ then $output \gets 8 + source$
		\newline
		else $output \gets 8$
	\end{quote}
\end{example}
\noindent
%In the example above, 
Assume $source$ to be a non-negative 32-bits integer which is uniformly distributed on that domain, 
then there are 16 possible values of $output$, from 8 to 23. 
Observing any number between 9 and 23 as an output reveals everything about the confidential $source$, 
whilst observing 8 leaks small information; there are many possible values of $source$ 
($0, 16, 17, 18, \ldots, 2^{32}-1$) which produce 8 as the output. 
QIF is defined as the average of the leakage over all possible cases 
and fails to capture the above situation 
because we cannot distinguish \textit{vulnerable} and \textit{secure} cases if we take the average. 
Hence, as argued in \cite{Bi16}, a notion for dynamic leakage should reflect 
individual leakage caused by observing an output. 
%In \cite{CHS19}, we proposed two new notions ($\QIFone$ and $\QIFtwo$) for dynamic leakage, 
%proved it computationally hard to calculate either $\QIFone$ or $\QIFtwo$ in general, 
%and also proposed a calculation method based on model counting, 
%which works efficiently in many practical cases. 

\begin{figure}[h]
	\centering	
	\includegraphics[scale=0.55]{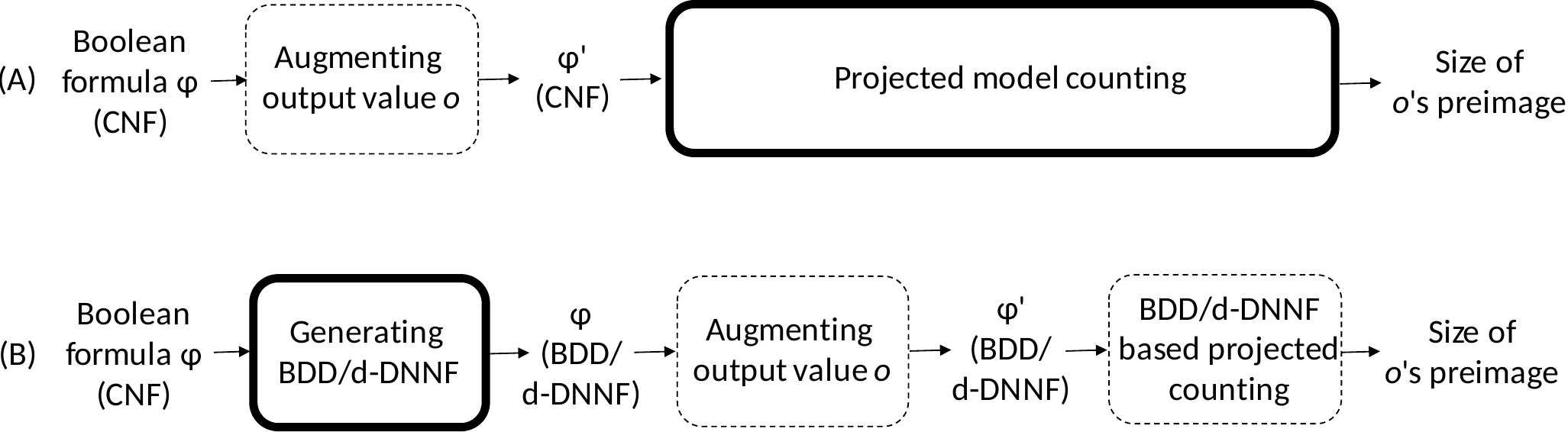}
	\caption{(A): Compute-on-Demand, (B): Construct-in-Advance.}
	\label{fig:1}
\end{figure}

As illustrated in Figure \ref{fig:1}, there are two different scenarios of quantifying dynamic leakage. 
We call the first scenario, which corresponds to diagram (A), 
\textit{\textbf{Compute-on-Demand}} (\textit{CoD}), 
and the second, which corresponds to diagram (B), \textit{\textbf{Construct-in-Advance}} (\textit{CiA}).
%
%\noindent
%In both (A) and (B), 
A box surrounded by bold lines represents a heavy-weighted process, which requires much computing resource. 
The main difference between (A) and (B) is the relative position of the heavy-weighted process, i.e., 
whether (A) to put the process after augmenting an observed output 
and then run the process each time we need (on demand) to compute dynamic leakage, 
or (B) to put the process before augmenting an observed output (in advance) so that 
we should run the process only once for one program. %
In \textit{CoD}, the heavy-weighted process is a projected model counting, 
for which off-the-shelf tools such as SharpCDCL \cite{SharpCDCL}, DSharp-p \cite{DSharp-p} and GPMC \cite{GPMC}
can be used. 
In \textit{CiA}, the heavy-weighted process is the one that generates BDD \cite{So99} or d-DNNF \cite{D01}, 
which are data structures to represent Boolean formulas. 
Generally, it takes time to generate BDD or d-DNNF but counting all solutions (models) by using them is easy. 
\textit{CiA} takes the full advantage of this characteristic.
Consider again Example \ref{ex:1.1} above. 
The set of all feasible pairs of ($source$, $output$) is $2^{32}$.  
%$\{(0,8),(1,9),(2,10),\ldots,(15,23),(16,8),(17,8),\ldots,(2^{32}-1,8) \}$, 
%of which cardinality is $2^{32}$. 
%Despite this is a very simple program, with such a big number of pairs, 
Even for such a simple program, 
using BDD or d-DNNF to store all those pairs is quite daunting in terms of both memory space and speed. 
Therefore, for programs with simple structure but a large number of input and output pairs, 
\textit{CoD} works better than \textit{CiA}. 
On the other hand, 
\textit{CiA} is preferable to \textit{CoD} when quantifying dynamic leakage 
is required many times on the same program. 
However, \textit{CoD} or \textit{CiA} alone is not a solution to the problem of scalability. 
\medskip\par
%we develop and extend their model to that of $\QIFone$ and $\QIFtwo$. 
In this paper, we introduce two compositional methods for computing dynamic leakage
inspired by the work of Kawamoto et al. \cite{KCP17} on the compositionality of static leakage. 
One is to utilize the sequential structure of a given program $P = P_1 ; P_2$. 
We first analyze $P_2$ and then compute the leakage of $P$ by 
analyzing $P_1$ based on the result on $P_2$. 
For the sequential composition, beside the benign yet time-consuming exact counting based on Breadth-First-Search (BFS), 
we also investigate an approximated approach. 
For upper bound of the count, we leverage the results on each sub-program by Max\#SAT in \cite{FRS17}. 
For lower bound of the count, we simply use Depth-First-Search (DFS) with timeout, i.e., 
DFS will stop when the execution time exceeds the predetermined timeout.

The other method we propose is based on the decomposition of the value domain of a program. 
For example, we divide the input domain as $I=I_1\cup I_2$ and the output domain
as $O=O_1\cup O_2$ of a program $P(I,O)$, 
compute the leakages for $P(I_i,O_j)$ for $i=1,2$ and $j=1,2$,
and then compute the leakage of the whole program $P(I,O)$ from them. 
This value domain based decomposition has two merits. 
First, it is flexible yet simple to adjust the components.
Secondly, the exact dynamic leakage of the composed program can be simply derived by taking the sum of those of its components. 
Despite the number of components can be large, this approach is promising with parallel computing. 
%
%the parallel and sequential compositions 
%as two basic building blocks to reason about the compositionality of dynamic leakage. 
%As sketched in Figure \ref{fig:2}, the parallel composition 
%of two programs has its input $\textbf{I}$ made from inputs $I_1$, $I_2$, 
%and output $\textbf{O}$ made from outputs $O_1$, $O_2$ of its components, 
%The sequential composition of two program has its input made from only input $I_1$ of the first component, 
%and its output made from only output $O_2$ of the second component. 
%Also for the sequential composition, input $I_2$ of $P_2$ is derived from $O_1$ of $P_1$. 
%\begin{figure}[h]
%	\centering
%	\includegraphics[scale=0.55]{fig2.eps}
%	\caption{Intuitive illustration of two basic types of composition.}
%	\label{fig:2}
%\end{figure}
%In general case of the parallel composition, based on Kawamoto et al., 
%we derive theoretical upper and lower bounds of the dynamic leakage of the container program from those of its components. 
%Go beyond that result, we propose a different type of parallel composition 
%based on combining separated value domains of shared input among components. 
%
\par
In summary, the contributions of this research are four-fold:
\begin{itemize}
	\item We propose a compositional method for dynamic leakage computation 
        based on the sequential structure inside a given program and the composability 
        of the leakage of the whole program from those of subprograms. 
	\item We propose another compositional method based on value domains, 
        which is suitable for parallel computing.
	\item We propose an approximated approach where we upper bound the count using Max\#SAT problem 
        and lower bound the count by DFS with predetermined timeout.
	\item We prototype a tool that can do parallel computation based on value domain decomposition
        and both exact counting and approximated counting for the sequential composition. 
        By using the tool, 
        we investigate feasibility and advantages of the proposed compositional methods for computing 
        dynamic leakage of several examples. 
\end{itemize}
\medskip
\par\noindent
{\bf Related work} 
\textit{Definitions of QIF:} 
%Despite the need to have a metric for information leakage is widely known, 
%There exists several definitions of QIF. 
Smith \cite{Sm09} gives a comprehensive summary on entropy-based QIF, 
such as Shannon entropy, guessing entropy and min entropy and compares them in various scenarios. 
Clarkson et al. \cite{CMS09}, on the other hand, include attacker's belief into their model. 
%Attempting to unify those definitions, 
Alvim et al. introduce gain function to generalize information leakage by separating
the probability distribution and the impact of individual information. 
\textit{Computational Complexity:} 
%Yasuoka and Terauchi \cite{YT10} proved that 
%the QIF comparison problem, which is not harder than the QIF calculation, 
%is not a k-safety property for any k.  
%Consequently, self-composition, a successful technique to verify non-interference property, 
%is no longer applicable to QIF. 
%Their subsequent work \cite{YT11} proves a similar hardness result for bounding QIF, 
Yasuoka and Terauchi \cite{YT11} prove complexity on computing QIF, 
including $PP$-hardness of precisely quantifying QIF for loop-free Boolean programs. 
Chadha and Ummels \cite{CU12} show that the QIF bounding problem of recursive Boolean programs 
is EXPTIME-complete. 
\textit{Precise Calculation:} 
In \cite{KMM13}, Klebanov et al. reduce QIF calculation to \#SAT problem projected on a specific set of variables. 
On the other hand, Phan et al. \cite{PM15} reduce QIF calculation to $\sharp$SMT problem 
to leverage existing SMT (satisfiability modulo theory) solvers. 
Recently, Val et al. \cite{VEBAH16} reported 
a SAT-based method that can scale to programs of 10,000 lines of code. 
% but still based on SAT solver and symbolic execution. 
\textit{Approximated Calculation:} 
%While precisely calculating QIF faces inherent hardness which prevent them from scaling up to practical systems, 
Approximation is a reasonable alternative for scalability. 
K\"{o}pf and Rybalchenko \cite{KR10} propose approximated QIF computation
by sandwiching the precise QIF with lower and upper bounds using randomization and abstraction, respectively, 
with a provable confidence. 
LeakWatch of Chothia et al. \cite{CKN14}, also give approximation 
with provable confidence by executing a program multiple times. 
Its descendant called HyLeak \cite{BKLT17} combines the randomization strategy of its ancestor with precise analysis. 
%Also using sampling but in Markov Chain Monte Carlo (MCMC) manner, 
Biondi et al. \cite{BEHLMQ18} utilize ApproxMC2, %
%an existing model counter created by some of the co-authors. 
%ApproxMC2 provides approximation on the number of models of a Boolean formula in CNF 
which provides approximation on the number of models of a Boolean formula in CNF
by Markov Chain Monte Carlo method. 
%with tunable precision and confidence. 
%ApproxMC2 uses hashing technique to divide the solution space into smaller buckets 
%with almost equal number of elements, then count the models for only one bucket and multiply it by the number of buckets. 
\textit{Composition of QIF:} 
Another attempt to the scalability is to break the system down into smaller fragments. 
%This direction is investigated very recently by Kawamoto et al. \cite{KCP17}. 
In \cite{KCP17}, Kawamoto et al. introduce two parallel compositions: with distinct inputs and with shared inputs, 
and give theoretical bounds on the leakage of the main program using those of the constituted sub-programs. 
Though our research was motivated by \cite{KCP17}, 
%\cite{KCP17} considers information channels as their targets while ours works on computer programs. 
we focus on a sequential structure of a target program and 
a decomposition of the value domain of the program 
while \cite{KCP17} uses a parallel structure of the target. 
\textit{Dynamic Leakage:} 
%McCamant et al. \cite{ME08} consider QIF as network flow through programs and 
%propose a dynamic analysis method that can work with executable files. 
%In general, information flow in \cite{ME08} is not the dynamic leakage we are considering in this paper. 
Bielova (\cite{Bi16}) discusses the importance of dynamic leakage and argues that 
any well-known QIF notion is not appropriate as a notion for dynamic leakage. 
Recently, we proposed two notions for dynamic leakage, $\QIFone$ and $\QIFtwo$ 
and give some results on computational complexity 
as well as a quantifying method based on model counting \cite{CHS19}. 
%\medskip
\par
%\noindent
%%\par\noindent
%%{\bf Structure of the remaining parts} 
The rest of the paper is organized as follows. 
We will review the definition of dynamic leakage and assume our program model in Section 2. 
Section 3 is dedicated to a method for computing dynamic leakage 
based on the sequential composition and also propose approximation methods. 
Section 4 proposes a parallel computation method based on value domain decomposition. 
Section 5 evaluates the proposed compositional methods including the comparison of 
\textit{CiA} vs. \textit{CoD} and exact vs. approximated computation 
based on the experimental results. 
Then the paper is concluded in Section 6.

\section{Preliminaries}
\subsection{Dynamic leakage}
The standard notion for static quantitative information flow (QIF) is defined as 
the mutual information between random
variables $S$ for secret input and $O$ for observable output:
\begin{equation} 
\QIF = H(S) - H(S|O)
%\label{eq:QIF}
\end{equation}
where 
$H(S)$ is the entropy of $S$ and 
$H(S|O)$ is the expected value of $H(S|o)$, 
which is the conditional entropy of $S$ when observing an output $o$. 
Shannon entropy and min-entropy are often used as the definition of entropy, and 
in either case, $H(S)-H(S|O)\ge 0$ always holds by the definition.

In \cite{Bi16}, the author discusses the appropriateness of the 
existing measures for dynamic QIF and points out their drawbacks, 
especially, each of these measures may become negative. 
Hereafter, let ${\cal S}$ and ${\cal O}$ denote the finite sets of input values and output values, 
respectively. 

\medskip\par
Let $P$ be a program with secret input variable $S$ 
and observable output variable $O$. 
For notational convenience, we identify 
the names of program variables with the corresponding 
random variables. Throughout the paper, we assume that a program always terminates. 
The syntax and semantics of programs assumed in this paper will be given 
in the next section. 
For $s\in {\cal S}$ and $o\in {\cal O}$, let 
$p_{SO}(s,o)$, $p_{O|S}(o|s)$, $p_{S|O}(s|o)$, $p_S(s)$, $p_O(o)$ denote 
the joint probability of $s\in {\cal S}$ and $o\in {\cal O}$, 
the conditional probability of $o\in {\cal O}$ given $s\in {\cal S}$ (the likelihood), 
the conditional probability of $s\in {\cal S}$ given $o\in {\cal O}$ (the posterior probability), 
the marginal probability of $s\in {\cal S}$ (the prior probability) and 
the marginal probability of $o\in {\cal O}$, respectively.
We often omit the subscripts as $p(s,o)$ and $p(o|s)$ if they are clear from the context.
By definition, 
$p(s,o)  =  p(s|o)p(o)=p(o|s)p(s)$, 
$p(o)  =  \sum_{s\in {\cal S}} p(s,o)$, 
$p(s)  =  \sum_{o\in {\cal O}} p(s,o)$. 
%\begin{eqnarray}
%\label{eq:Bayes}
%p(s,o) & = & p(s|o)p(o)=p(o|s)p(s), \\
%p(o) & = & \sum_{s\in {\cal S}} p(s,o), \\
%p(s) & = & \sum_{o\in {\cal O}} p(s,o). 
%\end{eqnarray}
%\noindent

We assume that (the source code of) $P$ and 
the prior probability $p(s)$ ($s\in {\cal S}$) are known to an attacker.
For $o\in {\cal O}$, let $\pre_P(o) = \{ s\in {\cal S} \mid p(s|o) > 0 \}$, which is 
called the preimage of $o$ (by the program $P$). 

\begin{figure}[h]
	\centering
	\includegraphics[scale=0.4]{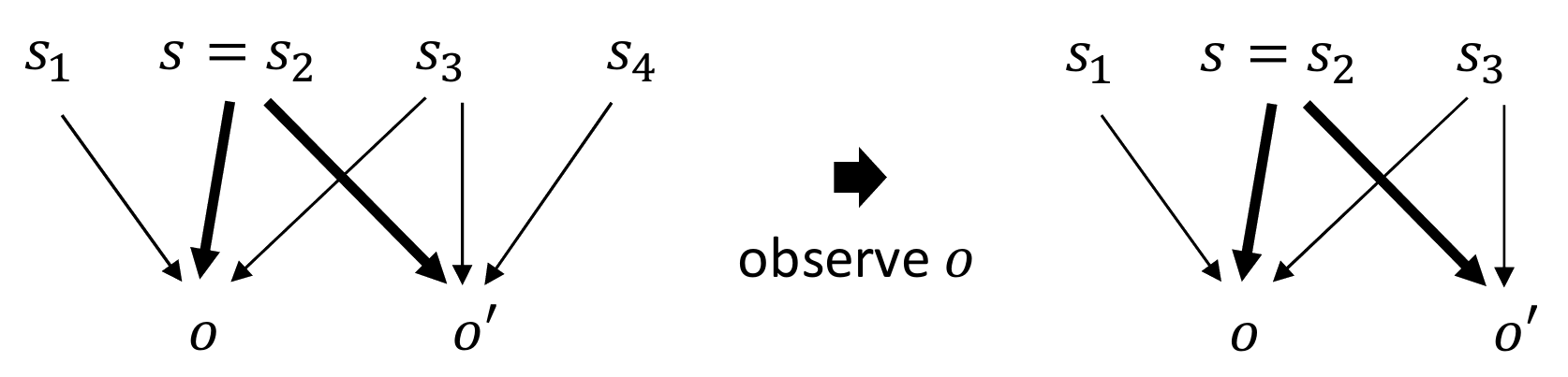}
	\includegraphics[scale=0.4]{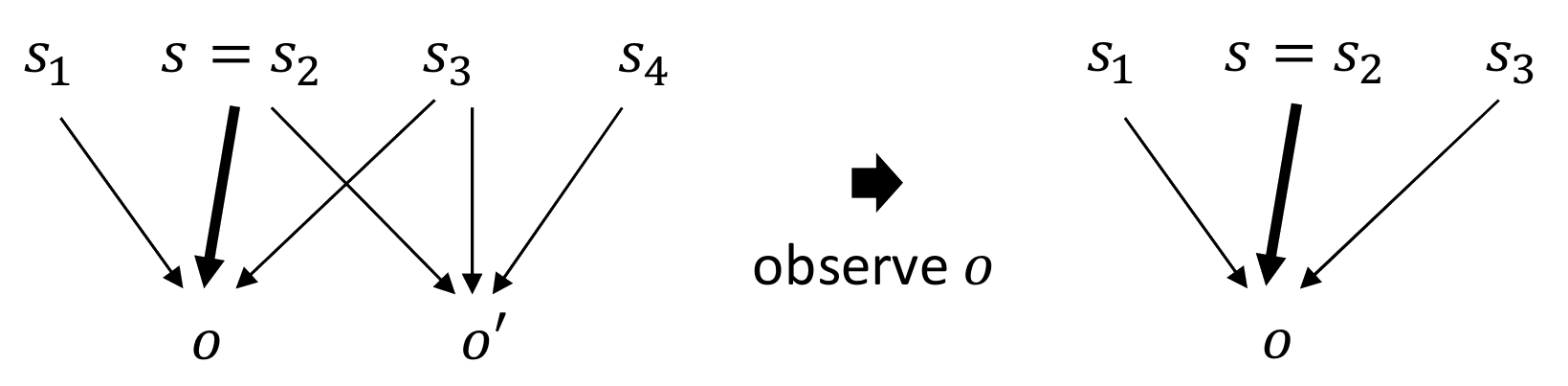}
	\caption{$\QIFone$ (the upper) and $\QIFtwo$ (the lower)}
	\label{fig:2.1}
\end{figure}

%We propose two notions for dynamic QIF
%when an attacker observes an output value $o\in {\cal O}$. 
%When we use Shannon entropy, both $H(S)$ and $H(S|O)$ are the expected value
%of the self-information and that of the conditional information given an output, respectively.
%In contrast, we are interested in the dynamic leakage when observing $o\in {\cal O}$. 
%Assume that the secret information is $s\in {\cal S}$. 
Considering the discussions in the literature, 
we define new notions for dynamic QIF that satisfy the following requirements \cite{CHS19}:
\begin{enumerate}
	\def\labelenumi{(R\arabic{enumi})}
	\def\theenumi{(R\arabic{enumi})}
	\item Dynamic QIF should be always non-negative because
	an attacker obtains some information (although sometimes very small or even zero) when 
	he observes an output of the program.
	\item It is desirable that dynamic QIF is independent of a secret input $s\in {\cal S}$. 
	Otherwise, the controller of the system may change the behavior for protection
	based on the estimated amount of the leakage that depends on $s$, 
	which may be a side channel for an attacker. 
	\item The new notion should be compatible with the existing notions when 
	we restrict ourselves to special cases such as deterministic programs, 
	uniformly distributed inputs, and taking the expected value. 
\end{enumerate}
%
%%%%%%%%%%%%%%% (AM1)
%%%%%%%%%%%%%%% 
%%%%%%%%%%%%%%% (Bm7)
The first notion is the self-information
of the secret inputs consistent with an observed output $o\in {\cal O}$.
Equivalently,
the attacker can narrow down the possible secret inputs
after observing $o$ to the preimage of $o$ by the program. 
We consider the self-information of $s \in \mathcal{S}$ after 
the observation as the probability of $s$ divided by the sum 
of the probabilities of the inputs in the preimage of $o$
(see the upper part of Fig. \ref{fig:2.1}).
\begin{eqnarray}
\QIFone^{P}(o) % & = & - \log p(s) + \log \frac{p(s)}{\sum_{p(s'|o)>0}p(s')}\\
& = & - \log (\sum_{s' \in \pre_P(o)} p(s')). \label{eq:QIF1}
\end{eqnarray}
The second notion is
the self-information of
the joint events $s'\in {\cal S}$ and an observed output $o\in {\cal O}$
(see the lower part of Fig. \ref{fig:2.1}).
%Since $p(o) = p(s,o)/p(s|o)$, this is equal to
This is equal to the the self-information of $o$.
%
%\begin{eqnarray}
%\QIFtwo^{P}(o) & = & - \log (\sum_{s'\in {\cal S}}p(s',o)) \\
%= - \log p(o) & = & - \log p(s,o) + \log p(s|o). \label{eq:QIF2}
%\end{eqnarray}
\begin{equation}
\QIFtwo^{P}(o) = - \log (\sum_{s'\in {\cal S}}p(s',o)) 
= - \log p(o) = - \log p(s,o) + \log p(s|o). \label{eq:QIF2}
\end{equation}
%%%%%%%%%%%%%%%
%
%%%%%%%%%%%%%%% (Bm5)
Both notions are defined by considering how much
possible secret input values are reduced by observing an output.
We propose two notions because there is a trade-off between
the easiness of calculation and the appropriateness \cite{CHS19}.
\begin{theorem}[\cite{CHS19}]
	\label{th:det-quant}
	If a program $P$ is deterministic, 
	for every $o\in{\cal O}$ and $s\in{\cal S}$, 
	\[
	\QIFone^{P}(o) = \QIFtwo^{P}(o) = - \log p(o).
	\]
	If input values are uniformly distributed, $\QIFone^{P}(o) = $
	$\log \frac{|{\cal S}|}{|\pre_P(o)|}$ for every $o\in {\cal O}$. 
	\qed
\end{theorem}
\subsection{Program model}
%Let $\Bool =\{ \False, \True \}$ be the set of truth values and 
%let $\Natz=\Nat\cup\{0\}$ be the set of natural numbers including 0 where
%$\Nat = \{1, 2, \ldots\}$.
%Also let $\Rat$ denote the set of rational numbers. 
We assume probabilistic programs where every variable stores a natural number
and the syntactical constructs are 
assignment to a variable, conditional, probabilistic choice, 
while loop and concatenation: 
%(see Table \ref{tbl:syntax}). 
%\begin{table}[h]
%\caption{Program syntax}
\begin{eqnarray*}
	b & ::= & \False \mid \True \mid \neg b \mid b\vee b \mid e < e \\
	e & ::= & X \mid n  \mid e + e \\
	c & ::= & \mbox{skip}
	\mid X \gets e 
	\mid \mbox{if } b \mbox{ then } c \mbox{ else } c \mbox{ end}\\
	& & 
	\mid c~_{r}[]_{1-r}~c 
	\mid \mbox{while } b \mbox{ do } c \mbox{ end}
	\mid c ; c
\end{eqnarray*}
where 
$<, X, n, +$ stand for 
a binary relation on natural numbers, 
a program variable, 
a constant natural number and
a binary operation on natural numbers, 
respectively, and 
$r$ is a constant rational number representing the branching probability for a choice
command where $0\le r\le 1$. 
In the above BNFs, objects derived from the syntactical categories $b$, $e$ and $c$
are called conditions, expressions and commands, respectively.
%\end{table}
A command $X \gets e$ assigns the value of expression $e$ to variable $X$. 
A command $c_{1~r}[]_{1-r}~c_2$ means that the program chooses $c_1$ with probability $r$
and $c_2$ with probability $1-r$.  
Note that this is the only probabilistic command. 
The semantics of the other constructs are defined in the usual way. 

A program $P$ has the following syntax:
\begin{eqnarray*}
	P & ::= & \mbox{in } \vec{S}; \mbox{ out } \vec{O}; \mbox{ local } \vec{Z}; c 
	\mid P ; P
\end{eqnarray*}
where $\vec{S}, \vec{O}, \vec{Z}$ are sequences of variables which are disjoint from one another. 
A program is required to satisfy the following constraints on variables. 
We first define $In(P), Out(P), Local(P)$ for a program $P$ as follows. 
\begin{itemize}
	\item 
	If $P = \mbox{in } \vec{S}; \mbox{ out } \vec{O}; \mbox{ local } \vec{Z}; c$, we define
	$In(P) = \{ V \mid V \mbox{ appears in } \vec{S} \}$, 
	$Out(P) = \{ V \mid V \mbox{ appears in } \vec{O} \}$ and $Local(P) = \{V \mid V \mbox{ appears in } \vec{Z} \}$. 
	In this case, we say $P$ is a simple program. 
	We require that no varible in $In(P)$ appears in the left-hand side of an assignment
	command in $P$, i.e., any input variable is not updated. 
	
	%
	%\item 
	%If $P = P_1 || P_2$, we define
	%$In(P) = In(P_1) \cup In(P_2)$, 
	%$Out(P) = Out(P_1) \cup Out(P_2)$ and $Local(P) = Local(P_1) \cup Local(P_2)$ where we require that $In(P_1)\cap In(P_2) = Out(P_1)\cap Out(P_2) = Local(P_1)\cap Local(P_2) = \emptyset$ holds.
	
	\item 
	If $P = P_1 ; P_2$, we define
	$In(P) = In(P_1)$, 
	$Out(P) = Out(P_2)$ where we require that 
	$In(P_2) = Out(P_1)$ holds. We also define $Local(P) = Local(P_1)\cup Local(P_2) \cup Out(P_1)$.
\end{itemize}
% 
%$Var(P) = \{ V \mid V \mbox{ appears in } \vec{X}, \vec{Y} \mbox{ or } \vec{Z} \}$. 
%We will use the same notation $Var(e)$ and $Var(c)$ for an expression $e$ and a command $c$. 
A program $P$ is also written as $P(S,O)$ 
where $S$ and $O$ are enumerations of $In(P)$ and $Out(P)$, respectively. 
%to emphasize the input and output variables 
%$\vec{S}$ and $\vec{O}$ of 
%$\pi_1 = \mbox{in } \vec{S}; \mbox{ out } \vec{O}; \mbox{ local } \vec{Z}; c.$
%
%A program $P_1 || P_2$ represents the parallel composition of $P_1$ and $P_2$ and 
A program $P_1 ; P_2$ represents the sequential composition of $P_1$ and $P_2$. 
Note that the semantics of $P_1 ; P_2$ is defined in the same way as that of 
the concatenation of commands $c_1 ; c_2$ except that 
the input and output variables are not always shared by $P_1$ and $P_2$ 
in the sequential composition. 
%
%For a program $P(\vec{S},\vec{O})$, 
%we let ${\cal S}, {\cal O}$ denote 
%the sets of input values and output values, respectively. 
%
%Hereafter, we abbreviate a sequence of variables $\vec{X}$ as $X$ if no confusion occurs. 
%So, we write $P(S,O)$ instead of $P(\vec{S},\vec{O})$ to mention a program $P$.
%
%
%A main procedure has the following syntax:
%
%\[ \mbox{in } \vec{S} \mbox{ out } \vec{O} \mbox{ local } \vec{Z}; c \]
%
%where $\vec{S}, \vec{O}, \vec{Z}$ are sequences of input, output and local variables, 
%respectively. 
%
%The size of a program $P$ is 
%the sum of the number of commands and 
%the number of variables in $P$. 
%
%If a program does not have a recursive procedure call and $k=1$, it is called a (non-recursive) while program. 
%If a while program does not have a while loop, it is called a loop-free program 
%If a program does not have a while loop, it is called a loop-free program 
%(or straight-line program). 
If a program does not have a probabilistic choice, it is {\em deterministic}.

\section{Sequential composition}
This section proposes a method of computing both exact and approximated dynamic leakage 
by using sequential composition. 
While the formula in exact calculation can be used for both probabilistic and deterministic programs, 
we consider only deterministic programs with uniformly distributed input.
\subsection{Exact calculation}

For a program $P(S,O)$, 
an input value $s\in {\cal S}$ and 
a subset ${\cal S}'$ of input values, let
\begin{eqnarray*}
	\Post_P(s)         & = & \{ o \mid p(o|s) > 0 \},\\
	\Post_P({\cal S}') & = & \bigcup_{s\in{\cal S}'} \Post_P(s).
\end{eqnarray*}
If $P$ is deterministic and $\Post_P(s) = \{ o \}$, we write $\Post_P(s) = o$. 

Let $P = P_1 ; P_2$ be a program. 
We assume that $In(P_1), Out(P_1), In(P_2), Out(P_2)$ are all singleton sets for simplicity. 
This assumption does not lose generality; for example, if $In(P_1)$ contains more than one variables, 
we instead introduce a new input variable that stores the tuple consisting of a value of each variable in $In(P_1)$. 
Let 
$In(P) = In(P_1) = \{ S \}$, 
$Out(P_1) = In(P_2) = \{ T \}$, 
$Out(P) = Out(P_2) = \{ O \}$, and 
let ${\cal S}, {\cal T}, {\cal O}$ be
the corresponding sets of values, respectively. 
%
%\[
%P(s) = t\gets P_1(s) ; P_2(t). 
%\]
%
For a given $o\in {\cal O}$, 
$\Pre_P(o)$ and $p(o)$, 
which are needed to compute $\QIFone^{P}(o)$ and $\QIFtwo^{P}(o)$, (see (\ref{eq:QIF1}) and (\ref{eq:QIF2}))
can be represented 
in terms of those of $P_1$ and $P_2$ as follows.

\begin{eqnarray}
\Pre_P(o) & = & \bigcup_{t\in(\Pre_{P_2}(o)\cap\Post_{P_1}({\cal S}))}\Pre_{P_1}(t),
\label{eq:sc1}\\
p(o) & = & \sum_{s\in {\cal S}, t\in {\cal T}} p(s) p_1(t|s) p_2(o|t).
\label{eq:sc2}
\end{eqnarray}
%
%Though the above equations give a compositional way of computing the exact values of 
%$\QIFone(o)$ and $\QIFtwo(o)$ for a given $o\in {\cal O}$, 
%the computational cost is not small in general.
%For example, if we use (\ref{eq:sc1}), 
%we have to enumerate $\Pre_{P_1}(t)$ for every $t\in (\Pre_{P_2}(o)\cap\Post_{P_1}({\cal S}))$.
If $p(s)$ is given, 
we can compute (\ref{eq:sc1}) by enumerating the sets $\cal T$ and $\Pre_{P_1}(t)$
for $t\in(\Pre_{P_2}(o)\cap\Post_{P_1}({\cal S}))$ and 
also for (\ref{eq:sc2}).
This approach can easily be generalized to the sequential composition of more than two programs, 
in which the enumeration is proceeded in Breadth-First-Search fashion. However, in this approach, 
search space will often explode rapidly and lose the advantage of composition. Therefore we come up with 
approximation, which is explained in the next subsection, as an alternative.

\subsection{Approximation}
Let us assume that $P(S,O)$ is deterministic and $S$ is uniformly distributed. In this subsection, we will 
derive both upper-bound and lower-bound of $|\Pre_P(o)|$ which provides lower-bound and upper-bound 
of $\QIFone^{P}(o) = \QIFtwo^{P}(o)$ respectively. In general, our method can be applied to the sequential composition of 
more than two sub-programs.
\vspace{-1em}
\subsubsection{Lower bound}
To infer a lower bound of $|\Pre_P(o)|$, we leverage Depth-First-Search (DFS) with a predefined timeout 
such that the algorithm will stop 
when the execution time exceeds the timeout and output the current result as the lower bound. 
The method is illustrated in Algorithm \ref{alg:lowerbound}.
The problem is defined as: 
given a program $P=P_1;P_2;\cdots;P_n$, an observable output $o$ of the last sub-program $P_n$ and a predetermined $timeout$, 
derive a lower bound of $|\Pre_{P}(o)|$ by those $n$ sub-programs.
\begin{algorithm}
\caption{LowerBound($P_1,\cdots,P_n, o$, timeout)}
\label{alg:lowerbound}
\begin{algorithmic}[1]
	\State $Pre[2..n] \gets empty$
	\State $Stack \gets empty$
	\State $level \gets n$
	\State $acc\_count \gets 0$
	\State Push($Stack, o$)
	\State $Pre[n] \gets EnumeratePre(P_n, o)$
	\While{not $Stack.empty$ and $execution\_time < timeout$}
		\If {$level = 1$}
			\State $acc\_count \gets acc\_count + $ CountPre($P_1, Stack.top$)
			\State $level \gets level + 1$
			\State Pop($Stack$)
		\Else 
			\State $v \gets $ PickNotSelected($Pre[level]$)
			\If {$v = AllSelected$}
				\State $level \gets level + 1$
				\State Pop($Stack$)
			\Else
				\State Push($Stack, v$)
				\State $level \gets level - 1$
				\If {$level > 1$}
					\State $Pre[level] \gets EnumeratePre(P_{level}, v)$
				\EndIf
			\EndIf
		\EndIf
	\EndWhile
	\State \Return $acc\_count$
\end{algorithmic}
\end{algorithm}

In Algorithm \ref{alg:lowerbound}, 
\textit{CountPre(Q, o)} counts $|\Pre_{Q}(o)|$, 
\textit{PickNotSelected(Pre[i])} select an element of $Pre[i]$ that has not been traversed yet or 
returns \textit{AllSelected} if there is no such element, and 
\textit{EnumeratePre($P_{i},v$)} lists all elements in $\Pre_{P_i}(v)$.
$Pre[i]$ stores $\Pre_{P_i}(o_i)$ for some $o_i$. 
%This array is maintained during the searching process. 
For $P_1$, it is not necessary to store its preimage because we need only the size of the preimage.
Lines 1 to 5 are for initialization. 
%$Pre$, $Stack$, $level$ (the index $i=level$ of $P_i$), 
%$acc\_count$ (accumulated count of $|\Pre_{P}|$).
Line 6 enumerates $\Pre_{P_n}(o)$. 
Lines 7 to 21 constitute the main loop of the algorithm, which is stopped 
either when the counting is done or when time is up. 
When $level=1$, lines 8 to 11 are executed and $CountPre$ will return $\Pre_{P_1}(Stack.top)$ in which $Stack.top$ is 
the input of $P_2$ that leads to output $o$ of $P_n$, then back-propagate; 
lines 13 to 16 check if all elements in the preimage set of the current level is already considered 
and if so, back-propagate, otherwise push the next element onto the top of $Stack$ and go to the next 
level. 
\begin{theorem}
	In Algorithm \ref{alg:lowerbound}, if $P_1,\cdots,P_n$ are deterministic, $acc\_count$, which is returned at line 22, 
	is a lower bound of the preimage size of $o$ by $P_1,\cdots,P_n$.
	\qed
\end{theorem} 
%\begin{algorithm}
%\caption{EnumeratePreWithHeuristic($P_2, P_1, v$)}
%\begin{algorithmic}[1]
%	\State $Pre \gets EnumeratePre(P2, v)$
%	\State $Pre \gets FilterOnebit(P1, Pre)$
%	\State \Return $Pre$
%\end{algorithmic}
%\end{algorithm}
\vspace{-2em}
\subsubsection{Upper bound}
For an upper bound of $|\Pre_P(o)|$ we use Max\#SAT problem \cite{FRS17}, which is defined as follows. 
\begin{defn}
	Given a propositional formula $\varphi(X,Y,Z)$ over sets of variables $X$, $Y$ and $Z$, the Max\#SAT problem 
	is to determine $max_X\#Y.\exists Z.\varphi(X,Y,Z)$.
\end{defn}
\noindent
Let us consider a program $Q$, $In(Q)$, $Out(Q)$ and $Local(Q)$ as $\varphi$, $Y$, $X$ and $Z$ respectively. 
Then, the solution $X$ to the Max\#SAT problem can be interpreted as the output value which 
has the biggest size of its preimage set. 
In other words, $|max_X\#Y.\exists Z.\varphi(X,Y,Z)|$ is an upper bound of $\Pre_{Q}$ over all feasible outputs. 
Therefore, the product of those upper bounds of $|\Pre_{P_i}|$ over all $i$($1\le i\le n$) is obviously 
an upper bound of $|\Pre_{P}|$. 
Algorithm \ref{alg:upperbound} computes this upper bound where 
$CountPre(P_n, o)$ returns the size of the preimage of $o$ by $P_n$; 
$MaxCount(P_i)$ computes the answer to the Max\#SAT problem for program $P_i$. 
We used the tool developed by the authors of \cite{FRS17}, 
which produces estimated bounds of Max\#SAT with tunable confidence and precision. 
As explained in \cite{FRS17}, 
the tool samples output values of $k$-fold self-composition of the original program. 
The greater $k$ is, the more precise the estimation is, 
but also the more complicated the calculation of each sampling is. Note that $MaxCount(P_i)$ can be computed in advance only once. 
%In other words, 
%given those values, Algorithm \ref{alg:upperbound} only computes the product of numbers.
\begin{algorithm}
	\caption{UpperBound($P_1,\cdots, P_n, o$)}
	\label{alg:upperbound}
	\begin{algorithmic}[1]
		\State $Result \gets CountPre(P_n, o)$
		\For {$i \gets 1$ to $n$}
			\State $Result \gets Result * MaxCount(P_i)$
		\EndFor
		\State \Return $Result$
	\end{algorithmic}
\end{algorithm}

\begin{theorem}
	In Algorithm \ref{alg:upperbound}, if $P_1,\cdots,P_n$ are deterministic, $Result$, which is returned at line 4, 
	is an upper bound of the preimage size of $o$ by $P_1,\cdots,P_n$.
	\qed
\end{theorem}

%Since a lower-bound of $|\Pre_P(o)|$ provides an upper-bound of $\QIFone(o)=\QIFtwo(o)$, 
%and an upper-bound of the dynamic leakage is a safe approximation of the exact value, 
%we will give two lower-bounds of $|\Pre_P(o)|$. 
%Later, we will show the results of approximated leakage computation 
%using these lower-bounds on some benchmark programs. 
%Let 
%\[
%\Pre_{P_2}(o) \cap \Post_{P_1}({\cal S}) = \{ t_1, \ldots, t_n \}
%\]
%where $|\Pre_{P_1}(t_{i-1})| \ge |\Pre_{P_1}(t_{i})|$ for every $i$ ($1\le i\le n$). 
%Since $P$ is deterministic, 
%$\Pre_{P_1}(t_i) \cap \Pre_{P_1}(t_j) = \emptyset$ for each $i,j$ 
%($i\not=j$, $1\le i,j\le n$), and 
%$\sum_{1\le j\le n}|\Pre_{P_1}(t_j)| = |\Pre_P(o)|$.
%Hence for any $c$ ($0\le c\le n$) as a cutting constant, 
%%
%\begin{equation}
%\sum_{1\le j\le c}|\Pre_{P_1}(t_j)| \le |\Pre_P(o)|. 
%\label{eq:lb1}
%\end{equation}
%%
%This technique is called a beam-search. 
%
%We have another obvious lower-bound of $|\Pre_P(o)|$ as follows.
%%
%\begin{equation}
%n\times\min_{j}\{|\Pre_{P_1}(t_j)| \mid 1\le j\le n \}\le |\Pre_P(o)|. 
%\label{eq:lb1}
%\end{equation}
%
\section{Value domain decomposition}

%In the previous subsection, we provide method of 
%computing $\QIFone$ and $\QIFtwo$ for a given program 
%along with the parallel structure of the program based on \cite{KCP17}. 
Another effective method for computing the dynamic leakage 
in a compositional way is to decompose the sets of 
input values and output values into several subsets, 
compute the leakage for the subprograms restricted to
those subsets, and compose the results to obtain 
the leakage of the whole program. 
The difference between the parallel composition in \cite{KCP17} and the proposed method 
is that in the former case, a program under analysis itself is divided into two subprograms that run in parallel,
and in the latter case, the computation of dynamic leakage is conducted in parallel by decomposing the sets of input 
and output values.

Let $P(S,O)$ be a program.
Assume that the sets of input values and output values, 
${\cal S}$ and ${\cal O}$, are decomposed into mutually disjoint subsets as
\begin{eqnarray*}
{\cal S} & = & {\cal S}_1 \uplus \cdots \uplus {\cal S}_k,\\
{\cal O} & = & {\cal O}_1 \uplus \cdots \uplus {\cal O}_l.
\end{eqnarray*}
For $1\le i\le k$ and $1\le j\le l$, 
let $P_{ij}$ be the program obtained from $P$ by restricting 
the set of input values to ${\cal S}_i$ and 
the set of output values to ${\cal O}_j$ where
if the output value $o$ of $P$ for an input value $s\in {\cal S}_i$ 
does not belong to ${\cal O}_j$, the output value of $P_{ij}$ for 
input $s$ is undefined. 

By definition, for a given $o\in {\cal O}_j$, 
\begin{equation}
%\QIFone(o) & = & - \log \sum_{s\in\bigcup_{1\le i\le k}\Pre_{P_{i,j}}(o)} \pi(s), \\
\Pre_P(o)  =  \bigcup_{1\le i\le k}\Pre_{P_{i,j}}(o).\tag{*}
\label{eq:decompose}
\end{equation}

By (\ref{eq:QIF1}) and (\ref{eq:QIF2}), we can compute $\QIFone$ and $\QIFtwo$
in a compositional way. 

By Theorem \ref{th:det-quant}, 
if $P$ is deterministic and the prior probability of $S$ is 
uniformly distributed, what we have to compute is $|\Pre_P(o)|$, which can be obtained by summing 
up each $|\Pre_{P_{i,j}}(o)|$ by (\ref{eq:decompose}).
%
%\begin{eqnarray*}
%|\Pre_P(o)| & = & \sum_{1\le i\le k}|\Pre_{P_{i,j}}(o)|.
%\end{eqnarray*}
\begin{eqnarray*}
	|\Pre_P(o)| & = & \sum_{1\le i\le k}|\Pre_{P_{i,j}}(o)|.
\end{eqnarray*}
\section{Experiments \protect\footnote{the benchmarks and prototype are public at: \newline
		bitbucket.org/trungchubao-nu/dla-composition/src/master/dla-composition/}}
This section will investigate answers for the following questions: 
(1) Is \textit{CiA} always better than \textit{CoD} or vice versa? 
(2) How can parallel computation based on the value domain decomposition improve the performance? 
(3) How does approximation in the sequential composition work in terms of precision and speed? 
We will examine those questions through a few examples.
\subsection{Setting up}
The experiments were conducted on %in the hardware of which specification is as following: 
Intel(R) Xeon(R) CPU ES-1620 v3 @ 3.5GHz x 8 (4 cores x 2 threads), 32GB RAM, CentOS Linux 7. 
For parallel computation, we use OpenMP \cite{DM98} library. % in a C program which is used as a wrapper of system calls.
At the very first phase to transform C programs into CNFs, we leveraged the well-known CBMC\cite{CBMC}. 
For the construction of a BDD from a CNF and the model counting and enumeration of the constructed BDD, 
we use an off-the-shell tool PC2BDD \cite{PC2BDD}. 
We use  PC2DDNNF \cite{PC2DDNNF} for the d-DNNF counterpart. 
Both of the tools are developed by one of the authors in another project. 
Besides, as an ordering of Boolean variables of a CNF greatly affects the BDD generation performance, 
we utilize FORCE \cite{AMS03} to optimize the ordering before transforming a CNF into a BDD. 
We use MaxCount \cite{MaxCount} for estimating the answer of Max\#SAT problem. 
We implemented a tool for Algorithms \ref{alg:lowerbound} and \ref{alg:upperbound} as well as 
the exact count in sequential compositions in Java.

\subsection{The grade protocol}
This benchmark is taken from \cite{P15}. 
By this experiment, we investigated answers for questions (1) and (2) mentioned at 
the beginning of this section. 
This benchmark sums up (then takes the average of) the grades of a group of students 
without revealing the grade of each student. 
We used the benchmark with \textit{4 students} and \textit{5 grades}, and all 
variables are of 16 bits. 
For model counting, we suppose the observed output (the sum of students' grades) 
to be \textbf{1}, and hence the number of models is 4. 
GPMC \cite{GPMC}, one of the fastest tools for quantifying dynamic leakage as shown in \cite{CHS19}, 
was chosen as the representative tool for \textit{CoD} approach.
We manually decompose the original program into 4, 8 and 32 sub-programs 
by adding constraints on input and output of the program based on the value domain decomposition
(the set of output values is divided into 2 and the set of input values is divided into 2, 4 or 16).
Table \ref{tbl_grade_protocol} is divided into sub-divisions corresponding to specific tasks: 
BDD construction, d-DNNF construction and model counting based on different approaches. 
In each sub-division, the \textbf{bold number} represents the shortest execution time in each column 
(i.e., the same number of decomposed sub-programs, but different numbers of threads) 
and the \textbf{\underline{underlined one}} represents the best in that sub-division. 
`$-$' represents cases when the number of threads is greater than the number of sub-programs, 
which are obviously meaningless to do experiments. 
\vspace{-1em}
\begin{table}[H]
\centering
\begin{tabular}{ |c|c|c|c|c|c|}
	\hline
	\multicolumn{2}{|c|}{} & $n=32$ & $n=8$ & $n=4$ & $n=1$\\
	\hline
	\multirow{5}{11em}{ \centering BDD Construction} & $t=32$ & $\textbf{218.53s}$ & $-$ & $-$ & $-$\\
	\cline{2-6}
	&$t=16$ & $222.27s$ & $-$ & $-$ & $-$ \\
	\cline{2-6}
	&$t=8$ & $237.54s$ & $\textbf{\underline{137.74s}}$ & $-$ & $-$ \\
	\cline{2-6}
	&$t=4$ & $254.88s$ & $144.55s$ & $\textbf{155.90s}$ & $-$ \\
	\cline{2-6}
	&$t=2$ & $376.21s$ & $233.34s$ & $214.65s$ & $-$ \\
	\cline{2-6}
	&$t=1$ & $736.74s$ & $450.85s$ & $391.99s$ & $\textbf{243.85}s$ \\
	\hline
	\multirow{5}{11em}{\centering d-DNNF Construction} & $t=32$ & $93.17s$ & $-$ & $-$ & $-$ \\
	\cline{2-6}
	&$t=16$ & $\textbf{\underline{91.49s}}$ & $-$ & $-$ & $-$ \\
	\cline{2-6}
	&$t=8$ & $107.31s$ & $\textbf{123.48s}$ & $-$ & $-$ \\
	\cline{2-6}
	&$t=4$ & $141.27s$ & $147.79s$ & $\textbf{175.34s}$ & $-$ \\
	\cline{2-6}
	&$t=2$ & $215.92s$ & $226.93s$ & $247.45s$ & $-$ \\
	\cline{2-6}
	&$t=1$ & $398.99s$ & $391.67s$ & $457.38s$ & $\textbf{304.88s}$ \\
	\hline
	\multirow{5}{11em}{ \centering Model Counting \newline (\textit{CiA} - BDD based)} & $t=32$ & $\textbf{0.21s}$ & $-$ & $-$ & $-$ \\
	\cline{2-6}
	&$t=16$ & $0.22s$ & $-$ & $-$ & $-$ \\
	\cline{2-6}
	&$t=8$ & $0.25s$ & $\textbf{\underline{0.13s}}$ & $-$ & $-$ \\
	\cline{2-6}
	&$t=4$ & $0.30s$ & $0.16s$ & $\textbf{0.16s}$ & $-$ \\
	\cline{2-6}
	&$t=2$ & $0.65s$ & $0.31s$ & $0.24s$ & $-$ \\
	\cline{2-6}
	&$t=1$ & $0.86s$ & $0.36s$ & $0.31s$ & $\textbf{0.30s}$ \\
	\hline
	\multirow{5}{11em}{ \centering Model Counting \newline (\textit{CiA} - d-DNNF based)} & $t=32$ & $\textbf{0.05s}$ & $-$ & $-$ & $-$ \\
	\cline{2-6}
	&$t=16$ & $\textbf{0.05s}$ & $-$ & $-$ & $-$ \\
	\cline{2-6}
	&$t=8$ & $\textbf{0.05s}$ & $\textbf{\underline{0.01s}}$ & $-$ & $-$ \\
	\cline{2-6}
	&$t=4$ & $0.07s$ & $\textbf{\underline{0.01s}}$ & $\textbf{\underline{0.01s}}$ & $-$ \\
	\cline{2-6}
	&$t=2$ & $0.12s$ & $0.02s$ & $0.02s$ & $-$ \\
	\cline{2-6}
	&$t=1$ & $0.18s$ & $0.04s$ & $0.03s$ & $\textbf{0.25s}$ \\
	\hline
	\begin{tabular}{@{}c@{}} Model Counting \\ (\textit{CoD} - using GPMC) \end{tabular}
	& $t=1$ & $-$ & $-$ & $-$ & $\textbf{\underline{44.69s}}$\\
	\hline
\end{tabular}
\newline
\newline
\caption{Excecution time for the construction of data structures and model counting of different approaches. 
$n$: number of sub-programs decomposed from the original program; 
$t$: number of threads specified by \textit{num\_thread} compiling directive of OpenMP.}
\label{tbl_grade_protocol}
\end{table}
\vspace{-1em}
\noindent
Let us keep in mind that $n=1$ means non-decomposition, 
$t=1$ means a serial execution and the number of physical CPUs of the hardware is 8.
From Table \ref{tbl_grade_protocol}, we can infer the following conclusion:% with some confidence:
\begin{itemize}
	\item In general, increasing the number of threads 
        (up to the number of sub-programs) does improve the execution time 
        in both the construction of data structures and the model counting.
	\item When the number of sub-programs is close to the number of physical CPUs (8), 
        the execution time is among the best if not the best.
	\item In this example, \textit{CiA} shows a huge improvement over \textit{CoD}, 
        which is more than 4000 times 
	($\textbf{\underline{0.01s}}$ vs. $\textbf{\underline{44.69s}}$) with the best tuning of the former. 
        Of course, \textit{CiA} takes time in constructing data structures BDD or d-DNNF.
\end{itemize}
\noindent
The performance with d-DNNF is better than that with BDD in this example, 
but this seems due to the implementation of the tools. 

\lstdefinestyle{CStyle}{
	basicstyle=\footnotesize,
	breakatwhitespace=false, 
	frame=single,        
	breaklines=true,                 
	captionpos=b,                    
	keepspaces=true,                 
	numbers=left,                    
	numbersep=5pt,                  
	showspaces=false,                
	showstringspaces=false,
	showtabs=false,                  
	tabsize=2
}

%\lstset{style=CStyle}
%\begin{lstlisting}[language=C, caption=grade.c]
%unsigned int nondet_int();
%int main(void){	
% unsigned int S=4, G=5; // number of students and grades
% unsigned int n, output, sum=0, i=0, j=0, c;
% unsigned int numbers[S], a[S], h[S];	
% n = ((G-1)*S)+1;
% for (c=0; c<S; c++) h[c]=nondet_int()%G;
% for (c=0; c<S; c++) numbers[c]=nondet_int()%n;	
% while (i<S) {
%  j=0;
%  while (j<G) {
%   if (h[i]==j) a[i]=j+numbers[i]-numbers[(i+1)%S];
%   j=j+1;
%  }
%  i=i+1;
% }
% for (c=0; c<S; c++) sum+=a[c];
% output=sum%n;
%}
%\end{lstlisting}
\vspace{-1em}
\subsection{Bit shuffle and Population count}
\textit{population\_count} is the 16-bit version of the benchmark of the same name given in \cite{P15}. 
In this experiment, the original program is decomposed into three sub-programs 
in such a way that each sub-program performs one bit operation of the original. 
Inspired by \textit{population\_count}, we created the benchmark \textit{bit\_shuffle}, which consists 
of two steps: firstly it counts the number of bit-ones in a given secret number 
(by \textit{population\_count})\footnote{To increase 
	the preimage size by the first part, we took the count modulo 6.}, 
then it shuffles those bits to produce an output value. Though in Section 3, we suppose programs to be 
deterministic, when it comes to calculate $\QIFone$, the theory part works for probabilistic programs as well.
Hence, even \textit{bit\_shuffle} is probabilistic, conducting experiments on it is still valid.

This original program is divided into two sub-programs corresponding to the two steps. 
All the original programs and the decomposed sub-programs are provided in appendix \ref{appendix_benchmarks}.
As shown in Table \ref{tbl_seq_constr}, the construction time of BDD and d-DNNF 
for \textit{bit\_shuffle} was improved significantly 
(more than 100 times for BDD, 8 times for d-DNNF) by the decomposition 
while the improvement of \textit{population\_count} was not large. 
This is because in the former case two sub-programs are connected at a bottle-neck point, 
i.e., given a certain output, its preimage by the second program always has exactly one element, 
which is the number of bit-ones of that output, while in the latter case there is no such bottle-neck point. 
Besides, parallel computing is available for the decomposed sub-programs 
and the result is a little bit better. 
Probably, as the number of sub-programs increases, the effect of parallel computing would be larger.

For model counting, we let an output value 
be 3 (the number of models is 13110) for \textit{bit\_shuffle} 
and 7 (the number of models is 11440) for \textit{population\_count}. 
Table \ref{tbl_seq_mc} presents the execution times for model counting 
%in cases of 
%using GPMC, exact count based on BDD, d-DNNF, and approximation count based on BDD, d-DNNF, 
where the underlined \underline{numbers} are the exact counts, 
the bold \textbf{excution times} are the best results among approaches for the exact count of each benchmark 
and the italic \textit{data} are of approximated calculations. 
The execution times for the lower bounds are predetermined timeouts, which were designed to be 
1/2, 1/5 and 1/10 of the time needed by the exact count, followed by the time by \textit{CoD}. 
In \textit{bit\_shuffle} benchmark, lower bounds based on d-DNNF were not 
improved (all are zero) even the timeout was increased. 
This happened because an intermediate result of counting for one d-DNNF 
is unknown until the counting completes
while this benchmark contains only two sub-programs 
and the size of the preimage by the second sub-program is always one
(i.e., the number of times to count d-DNNFs is only two, one for the first sub-program and one for the second one).

\begin{table}[H]
	\centering
	\begin{tabular}{ |c|c|c|c|c|}
		\hline
		\multicolumn{2}{|c|}{}  & non-decompose & decompose (serial) & decompose (parallel)\\
		\hline
		\multirow{2}{6em}{BDD Construction} & bit\_shuffle & $>$1 hour & 33.90s & 33.46s\\
		\cline{2-5}
		& population\_count & 0.48s & 0.66s & 0.40s\\
		\hline
		\multirow{2}{6em}{d-DNNF Construction} & bit\_shuffle & 424.64s & 50.28s & 48.39s\\
		\cline{2-5}
		& population\_count & 1.19s & 0.71s & 0.69s\\
		\hline
	\end{tabular}
	\newline
	\newline
	\caption{BDD and d-DNNF construction time for different approaches.}
	\label{tbl_seq_constr}
\end{table}
\vspace{-2em}
\noindent

\begin{table}[H]
	\centering
		\begin{tabular}{ |c|c|c|c|c|c|c|}
			\hline
			\multicolumn{3}{|c|}{}&\multicolumn{2}{c}{\begin{tabular}{c}\multirow{2}{4em}{bit\_shuffle}\end{tabular}}&\multicolumn{2}{|c|}{\begin{tabular}{c}\multirow{2}{7.5em}{population\_count}\end{tabular}}\\[2ex]
			\hline
			\multicolumn{3}{|c|}{\textit{CoD} using GPMC} & 0.49s & \underline{13110} & \textbf{0.09s} & \underline{11440}\\
			\hline
			\multirow{6}{9em}{\textit{CiA}-BDD based}&\multicolumn{2}{|c|}{Exact count} & 1.47s & \underline{13110} & 10.98s & \underline{11440}\\
			\cline{2-7}
			& \multirow{5}{9em}{Approximation} & \multirow{4}{6em}{Lower bound} & \textit{0.75s} & \textit{6243} & \textit{5.5s} & \textit{5776}\\
			\cline{4-7}
			& & & \textit{0.30s} & \textit{1918} & \textit{2.2s} & \textit{888} \\
			\cline{4-7}
			& & & \textit{0.15s} & \textit{574} & \textit{1.1s} & \textit{312} \\
			\cline{4-7}
			& & & \textit{0.49s} & \textit{3713} & \textit{0.09s} & \textit{0} \\
			\cline{3-7}
			& & Upper bound & 0.02s & 14025 & 0.07s & 5898240\\
			\hline
			\multirow{6}{9em}{\textit{CiA}-d-DNNF based}&\multicolumn{2}{|c|}{Exact count} & \textbf{0.27s} & \underline{13110} & 3.50s & \underline{11440}\\
			\cline{2-7}
			& \multirow{5}{9em}{Approximation} & \multirow{4}{6em}{Lower bound} & \textit{0.13s} & \textit{0} & \textit{1.75s} & \textit{4712}\\
			\cline{4-7}
			& & & \textit{0.05s} & \textit{0} & \textit{0.70s} & \textit{1314} \\
			\cline{4-7}
			& & & \textit{0.03s} & \textit{0} & \textit{0.35s} & \textit{52} \\
			\cline{4-7}
			& & & \textit{0.49s} & \textit{13110} & \textit{0.09s} & \textit{0} \\
			\cline{3-7}
			& & Upper bound & 0.07s & 14025 & 0.13s & 5898240\\
			\hline
			
		\end{tabular}
		\newline
		\newline
		\caption{Model counting: execution time and the changing of approximation precision.}
		\label{tbl_seq_mc}
\end{table}
\noindent
 
From the experimental results, we obtain the following observations. %can infer the followings at some degree of confidence:
\begin{itemize}
	\item In the previous section (grade protocol) \textit{CiA} did much better than \textit{CoD} while
	     in this section, especially for \textit{population\_count} benchmark, \textit{CoD} using GPMC offered a huge improvement 
	     over \textit{CiA}. So the answer for question (1) is `No'. 
	\item While the precision of the upper bounds was not so good, 
        their execution times were small. 
        The precision could be improved by tuning the decomposition. 
        So this result could be a hint to a research on how to make a good decomposition to benefit the upper 
	bound approximation. 
	\item %The lower bounds were exactly bounded by the precise counting. 
        As expected, the lower bounds were improved as the timeout was lengthened. 
        Note that a lower bound of the model count corresponds to an upper bound of $\QIFone$. 
        Therefore, if we set a threshold for the leakage of a program, 
        we only need to know whether the lower bound of the counting exceedes 
        the constant corresponding to the threshold, 
        and if so, we can terminate the analysis but still be sure about the safety of the program.
\end{itemize}

\section{Conclusion}
In this paper, we focused on the efficient computation of dynamic leakage of a program
and considered two approaches \textit{Compute-on-Demand (CoD)} and \textit{Construct-in-Advance (CiA)}.
Then, we proposed two compositional methods, namely, 
computation along with the sequential structure of the program and 
parallel computation based on value domain decomposition. 
In the first method, we also proposed approximations that give both 
lower bound and upper bound of model counting. 
Our experimental result showed that: 
(1) both \textit{CiA} and \textit{CoD} are important because 
sometimes the former works better and the other times does the latter; 
(2) Parallel computation based on value domain decomposition works well generally; 
and (3) the precision of upper bound depends on the way of decomposition 
while that of lower bound depends on the preset timeout. 
However, all decomposition in the experiments were done manually and 
finding a systematical way of deciding a good decomposition is left as future work. 
Both BDD and d-DNNF have many applications other than computing dynamic leakage, 
but there is still a bottle neck at generating them from an object to be analyzed. 
One of the approaches in this paper 
\textit{composition based on value domains} can be a hint to speed up that process.

\appendix
\section{Benchmarks}\label{appendix_benchmarks}
\subsection{Grade protocol}
\subsubsection{Original program}
\lstset{style=CStyle}
\begin{lstlisting}[language=C, caption=grade.c]
typedef unsigned int size_t;
size_t nondet_int();
int main(void){	
size_t S = 4, G=5; //number of students, number of grades.
size_t n, output, sum = 0, i=0, j=0, c=0;
size_t numbers[S], announcements[S], h[S];

n = ((G-1)*S)+1;
for (c = 0; c < S; c++) { h[c] = nondet_int() % G;}
for (c = 0; c < S; c++) { numbers[c] = nondet_int() % n;}
while (i<S) {
	j=0;
	while (j<G) {
		if (h[i]==j)
		announcements[i]=j+numbers[i]-numbers[(i+1)%S];
		j=j+1;
	}
	i=i+1;
}

//computing the sum, producing the output and terminating
for (c = 0; c < S; c++) { sum += announcements[c]; }
output = sum % n;

assert(0);
return 0;
}
\end{lstlisting}
\subsubsection{One of sub-programs in 32-parts decomposition}
\lstset{style=CStyle}
\begin{lstlisting}[language=C, caption=grade\_32\_01.c]
typedef unsigned int size_t;
size_t nondet_int();
int main(void){	
size_t S = 4, G=5; //number of students, number of grades.
size_t n, output, sum = 0, i=0, j=0, c=0;
size_t numbers[S], announcements[S], h[S];

n = ((G-1)*S)+1;
for (c = 0; c < S; c++) { h[c] = nondet_int() % G;}
__CPROVER_assume (h[0]>=0 && h[0]<3);
__CPROVER_assume (h[1]>=0 && h[1]<3);
__CPROVER_assume (h[2]>=0 && h[2]<3);
__CPROVER_assume (h[3]>=0 && h[3]<3);
for (c = 0; c < S; c++) { numbers[c] = nondet_int() % n;}
while (i<S) {
j=0;
while (j<G) {
if (h[i]==j)
announcements[i]=j+numbers[i]-numbers[(i+1)%S];
j=j+1;
}
i=i+1;
}

//computing the sum, producing the output and terminating
for (c = 0; c < S; c++) { sum += announcements[c]; }
output = sum % n;

assert(output < 0 || output > 8);
return 0;
}
\end{lstlisting}

\subsection{Bit Shuffle}
\subsubsection{Original program}
\lstset{style=CStyle}
\begin{lstlisting}[language=C, caption=bit\_shuffle.c]
unsigned short int nondet_int();
int main(void){
unsigned short int s, s0, o=0, count;
s0=s;
s=(s&0x5555)+((s>>1)&0x5555);
s=(s&0x3333)+((s>>2)&0x3333);
s=(s&0x0f0f)+((s>>4)&0x0f0f);
count=(s+(s>>8))&0xff;
count=count%6;
unsigned short int bit_arr[16];
unsigned short int indices[16];

// initialize
for (unsigned int i=0; i<16; i++){
bit_arr[i]=0;
indices[i]=i;
}

// shuffle
for (unsigned int i=16; i>16-count; i--) {
unsigned int j = nondet_int()%i;
bit_arr[indices[j]]=1;
unsigned int temp=indices[j];
indices[j]=indices[i-1];
indices[i-1]=temp;
}

// generate result
for (unsigned int i=0; i<16; i++) {
if (bit_arr[i] == 1) o += 1 << (16 - i - 1);
}	
return 0;
}
\end{lstlisting}
\subsubsection{Sub-programs}
\lstset{style=CStyle}
\begin{lstlisting}[language=C, caption=bit\_shuffle\_2\_1.c]
typedef unsigned short int g_type;
g_type nondet_int();
int main(void) {
g_type s, s0;
g_type count;

s0 = s;
s = (s & 0x5555) + ((s >> 1) & 0x5555);
s = (s & 0x3333) + ((s >> 2) & 0x3333);
s = (s & 0x0f0f) + ((s >> 4) & 0x0f0f);
count = (s + (s>>8)) & 0xff;
count = count % 6;

assert(0);		
return 0;
}
\end{lstlisting}
\lstset{style=CStyle}
\begin{lstlisting}[language=C, caption=bit\_shuffle\_2\_2.c]
typedef unsigned short int g_type;
g_type nondet_int();
int main(void) {
g_type o, count;

count = count % 6;
unsigned short int bit_arr[16], indices[16];

// initialize
for (unsigned int i=0; i<16; i++) {
	bit_arr[i] = 0;
	indices[i] = i;
}

// shuffle
for (unsigned int i=16; i>16-count; i--) {
	unsigned int j = nondet_int() % i;
	bit_arr[indices[j]] = 1;
	// swap
	unsigned int temp = indices[j];
	indices[j] = indices[i-1];
	indices[i-1] = temp;
}

// generate result
o = 0;

for (unsigned int i=0; i<16; i++) { if (bit_arr[i] == 1) o += 1 << (16 - i - 1);}
assert(0);		
return 0;
}

\end{lstlisting}

\subsection{Population Count}
\subsubsection{Original programs}
\lstset{style=CStyle}
\begin{lstlisting}[language=C, caption=population\_count.c]
int main(void) {
unsigned short int S0, S, Output;
S0 = S;
S = (S & 0x5555) + ((S >> 1) & 0x5555);
S = (S & 0x3333) + ((S >> 2) & 0x3333);
S = (S & 0x0f0f) + ((S >> 4) & 0x0f0f);
Output = (S + (S>>8)) & 0xff;
assert(0);
return Output;
}
\end{lstlisting}
\subsubsection{Sub-programs}
\lstset{style=CStyle}
\begin{lstlisting}[language=C, caption=population\_count\_3\_1.c]
int main(void){
unsigned short int S0, S;
S0 = S;
S = (S & 0x5555) + ((S >> 1) & 0x5555);
assert(0);
return S;
}
\end{lstlisting}
\lstset{style=CStyle}
\begin{lstlisting}[language=C, caption=population\_count\_3\_2.c]
int main(void){
unsigned short int S0, S;
S0 = S;
S = (S & 0x3333) + ((S >> 2) & 0x3333);
assert(0);
return S;
}
\end{lstlisting}
\lstset{style=CStyle}
\begin{lstlisting}[language=C, caption=population\_count\_3\_3.c]
int main(void){
unsigned short int S0, S, Output;
S0 = S;
S = (S & 0x0f0f) + ((S >> 4) & 0x0f0f);
Output = (S + (S>>8)) & 0xff;
assert(0);
return Output;
}
\end{lstlisting}


\begin{thebibliography}{8}
\bibitem{AMS03}
F.\ Aloul, I.\ Markov, K.\ Sakallah,
\textit{FORCE: a fast and easy-to-implement variable-ordering heuristic},
Great Lakes Symposium on VLSI (GLSVLSI), 2003, 116--119.

\bibitem{ACPS12}
M.\ S.\ Alvim, K.\ Chatzikokolakis, C.\ Palamidessi, G.\ Smith,
\textit{Measuring information leakage using generalized gain functions},
21st Computer Security Foundations Symposium (CSF), 2012, 280--290.

\bibitem{Bi16}
N.\ Bielova,
\textit{Dynamic leakage - a need for a new quantitative information flow measure},
ACM Workshop on Programming Languages and Analysis for Security (PLAS), 2016, 83--88.

\bibitem{BEHLMQ18}
F.\ Biondi, M.\ A.\ Enescu, A.\ Heuser, A.\ Legay, K.\ S.\ Meel, J.\ Quilbeuf,
\textit{Scalable approximation of quantitative information flow in programs},
Verification, Model Checking, and Abstract Interpretation (VMCAI), 2018, 71--93.

\bibitem{BKLT17}
F.\ Biondi, Y.\ Kawamoto, A.\ Legay, L.\ M.\ Traonouez,
\textit{HyLeak: hybrid analysis tool for information leakage},
Automated Technology for Verification and Analysis (ATVA), 2017, 156--163.

\bibitem{CU12}
R.\ Chadha and M.\ Ummels,
\textit{The complexity of quantitative information flow in recursive programs},
Research Report LSV-2012-15, Laboratoire Sp\'{e}cification \& V\'{e}rification, \'{E}cole Normale Sup\'{e}rieure de Cachan, 2012.

\bibitem{CKN14}
T.\ Chothia, Y.\ Kawamoto, C.\ Novakovic,
\textit{LeakWatch: estimating information leakage from Java programs},
19th European Symposium on Research in Computer Security
(ESORICS), 2014, 219--236.

\bibitem{CHS19}
B.\ T.\ Chu, K.\ Hashimoto, H.\ Seki,
\textit{Quantifying dynamic leakage: complexity analysis and model counting-based calculation},
\url{https://arxiv.org/abs/1903.03802}.

\bibitem{CMS09}
M.\ R.\ Clarkson, A.\ C.\ Myers and F.\ B.\ Schneider,
\textit{Quantifying information flow with beliefs},
18th Computer Security Foundations Symposium (CSF), 2009, 655--701.

\bibitem{DM98}
L.\ Dagum, R.\ Menon,
\textit{OpenMP: an industry-standard API for shared-memory programming},
IEEE Computational Science \& Engineering, Volume 5 Issue 1, Jan 1998, 46--55.

\bibitem{D01}
A.\ Darwiche,
\textit{On the tractability of counting theory models and its application to belief revision and truth maintenance},
Jounal of Applied Non-Classical Logics 11(1-2), 2001, 11--34.

\bibitem{FRS17}
D.\ J.\ Fremont, M.\ N.\ Rabe, S.\ A.\ Seshia,
\textit{Maximum Model Counting},
AAAI Conference on Artificial Intelligence, 2017, 3885--3892.

\bibitem{GM82}
J.\ A.\ Goguen, J.\ Meseguer,
\textit{Security policies and security models},
IEEE Symposium on Security and Privacy (S$\&$P), 1982, 11--20.

\bibitem{KCP17}
Y.\ Kawamoto, K.\ Chatzikokolakis, C.\ Palamidessi,
\textit{On the compositionality of quantitative information flow},
Logical Methods in Computer Science, Vol. 13(3:11) 2017, pp. 1--31.

\bibitem{KMM13}
V.\ Klebanov, N.\ Manthey, C.\ Muise,
\textit{SAT-based analysis and quantification of information flow in programs},
Quantitative Evaluation of Systems (QEST), 2013, 177-192.

\bibitem{KR10}
B.\ K\"{o}pf, A.\ Rybalchenko,
\textit{Approximation and randomization for quantitative information flow analysis},
23rd Computer Security Foundations Symposium (CSF), 2010, 3--14.

\bibitem{P15}
Q.\ S.\ Phan,
\textit{Model counting modulo theories},
PhD thesis, Queen Mary University of London, 2015.

\bibitem{PM15}
Q.\ S.\ Phan, P.\ Malacaria,
\textit{All-solution satisfiability modulo theories: applications, algorithms and benchmarks},
10th International Conference on Availability, Reliability and Security (ARES), 2015, 100--109.

\bibitem{Sm09}
G.\ Smith,
\textit{On the foundations of quantitative information flow},
12th International Conference on Foundations of Software Science and Computational Structures (FOSSACS), 2009, 288--302.

\bibitem{So99}
F.\ Somenzi,
\textit{Binary decision diagrams},
\url{http://www.ecs.umass.edu/ece/labs/vlsicad/ece667/reading/somenzi99bdd.pdf}, 1999.

\bibitem{VEBAH16}
C.\ G.\ Val, M.\ A.\ Enescu, S.\ Bayless, W.\ Aiello, A.\ J.\ Hu,
\textit{Precisely measuring quantitative information flow: 10k lines of code and beyond},
IEEE European Symposium on Security and Privacy (EuroS$\&$P), 2016, 31--46.

\bibitem{YT11}
H.\ Yasuoka, T.\ Terauchi,
\textit{On bounding problems of quantitative information flow},
Journal of Computer Security (JCS), Vol. 19, 2011 November, 1029--1082.

%%% Tools

\bibitem{CBMC}
C Bounded Model Checker,
\url{https://www.cprover.org/cbmc}.

\bibitem{DSharp-p}
DSharp-p,
\url{https://formal.iti.kit.edu/~klebanov/software/}

\bibitem{GPMC}
GPMC,
\url{https://www.trs.css.i.nagoya-u.ac.jp/~k-hasimt/tools/gpmc.html}

\bibitem{MaxCount}
MaxCount,
\url{https://github.com/dfremont/maxcount}

\bibitem{PC2BDD}
PC2BDD,
\url{https://git.trs.css.i.nagoya-u.ac.jp/t_isogai/cnf2bdd}

\bibitem{PC2DDNNF}
PC2DDNNF,
\url{https://git.trs.css.i.nagoya-u.ac.jp/k-hasimt/gpmc-dnnf}

\bibitem{SharpCDCL}
SharpCDCL,
\url{http://tools.computational-logic.org/content/sharpCDCL.php}


%Author, F.: Article title. Journal \textbf{2}(5), 99--110 (2016)
%Author, F., Author, S.: Title of a proceedings paper. In: Editor,
%F., Editor, S. (eds.) CONFERENCE 2016, LNCS, vol. 9999, pp. 1--13.
%Springer, Heidelberg (2016). \doi{10.10007/1234567890}
%Author, F., Author, S., Author, T.: Book title. 2nd edn. Publisher,
%Author, A.-B.: Contribution title. In: 9th International Proceedings
%on Proceedings, pp. 1--2. Publisher, Location (2010)
%LNCS Homepage, \url{http://www.springer.com/lncs}. Last accessed 4
\end{thebibliography}
\end{document}